# RUM-NN: A Neural Network Model Compatible with Random Utility Maximisation for Discrete Choice Setups

Niousha Bagheri[1], Milad Ghasri[1, ✉], Michael Barlow[2]

1) School of Engineering and Technology, UNSW Canberra, ACT, 2600, Australia
2) School of Systems and Computing, UNSW Canberra, ACT, 2600, Australia

✉Corresponding author: m.ghasri@unsw.edu.au

## Abstract

This paper introduces a framework for capturing stochasticity of choice probabilities in neural networks, derived from and fully consistent with the Random Utility Maximization (RUM) theory, referred to as RUM-NN. Neural network models show remarkable performance compared with statistical models; however, they are often criticized for their lack of transparency and interpretability. The proposed RUM-NN is introduced in both linear and non-linear structures. The linear RUM-NN retains the interpretability and identifiability of traditional econometric discrete choice models while using neural network-based estimation techniques. The non-linear RUM-NN extends the model's flexibility and predictive capabilities to capture nonlinear relationships between variables within utility functions. Additionally, the RUM-NN allows for the implementation of various parametric distributions for unobserved error components in the utility function and captures correlations among error terms. The performance of RUM-NN in parameter recovery and prediction accuracy is rigorously evaluated using synthetic datasets through Monte Carlo experiments. Additionally, RUM-NN is evaluated on the Swissmetro and the London Passenger Mode Choice (LPMC) datasets with different sets of distribution assumptions for the error component. The results demonstrate that RUM-NN under linear utility structure and IID Gumbel error terms can replicate the performance of Multinomial Logit (MNL) model, but relaxing those constraints yields to superior performance for both Swissmetro and LMPC datasets. By introducing a novel estimation approach aligned with statistical theories, this study empowers econometricians to harness the advantages of neural network models.

**Keywords**: Neural Networks, Econometrics methods, Discrete choice modelling, Cholesky decomposition

## 1. Introduction

Discrete choice models are used to analyse and predict choices made from a set of discrete alternatives (Train, 2009). By quantifying the influence of various factors on the probability of each choice, it provides insights into individual preferences and trade-offs. These models have been widely used in various contexts, including travel mode choice (McFadden, 1973). This modelling is significant in understanding human choice behaviour as it allows for the prediction of decision patterns, and analysis of preferences, and supports the development of effective policies and marketing strategies (Bhat et al., 2007).

Choice modelling includes two primary paradigms of theory-driven and data-driven approaches. Theory-driven models provide interpretability by design, as they are built upon predefined theoretical foundations that dictate the relationships between dependent and independent variables. One such framework is models based on the Random Utility Maximisation (RUM) theory, which assumes that decision-makers select the alternative with the highest utility. Early discrete choice models adopted a linear structure for utility functions and made simplifying assumptions about the stochastic components of utility to achieve tractable specifications. The Multinomial Logit (MNL) model is the most comonly used example of this category (Train, 2009). Despite its simplicity, the MNL model is limited by the assumption that the error component follows a Gumbel distribution, which can be restrictive in practical applications. To enhance realism and improve predictive accuracy, more sophisticated models have been developed. Among these, the Multinomial Probit (MNP) assumes normally distributed error terms and accommodates correlations across choices (Train, 2009). On the other hand, data-driven models prioritize empirical evidence, utilising methods such as Machine Learning (ML) models to uncover patterns without strong theoretical assumptions about the decision-making processes. Data-driven models, such as neural networks, do not impose predefined behavioural theories on the relationship between dependent and independent variables, allowing the model to learn complex relationships from data. Neural networks, with their numerous parameters and high flexibility, often yield superior goodness-of-fit and prediction accuracy (Goodfellow et al., 2016).

Neural networks have emerged as a subset of Artificial Intelligence (AI) with widespread use across various fields such as image classification, speech recognition, natural language programming (Goodfellow et al., 2016), as well as transport planning and engineering (Cui et al., 2018, Zheng et al., 2021). Pekel and Kara (2017) reviewed studies on the application of neural network models in the context of transport and highlighted their prominence due to the opportunities they provide for accurate prediction, comparison, and evaluation in this field. However, the challenge with neural networks lies in their inherent complexity and lack of direct interpretability. Neural network models in real-case applications usually comprise hundreds and even thousands of parameters, making it challenging to directly explain the relationship between variables and the role of parameters (Zhao et al., 2020). Nonetheless, a wide range of interpretable techniques for neural networks known as Post-hoc explainability techniques are proposed to extract knowledge from these complex models (Arrieta et al., 2020). Almost all post-hoc explainability methods attempt to elucidate the behaviour of neural networks by subjecting them to a wide range of input scenarios. While these methods can shed light on model behaviour and gauge the sensitivity of output to different input variables, they do not fully address the lack of behavioural foundation in neural networks.

Recent studies have attempted to bridge this gap by introducing specific neural networks that align with behavioural theories e.g. (Sifringer et al., 2020, Wang et al., 2021, Wong and Farooq, 2021). These studies aim to blend theory-based knowledge and data-driven methods in neural networks. As an example, Bentz and Merunka (2000) introduced a feedforward Deep Neural Network (DNN) architecture with a Softmax activation function for the output layer, which resembles the probability function in an MNL specification. Similarly, Xia et al. (2023) introduced a Random Effect-Bayesian Neural Network (RE-BNN) framework that combines the Random Effect (RE) model with Bayesian Neural Networks (BNN). Similar to other neural networks designed to integrate behavioural theories, RE-BNN applies the Softmax function to calculate choice probabilities. The use of the Softmax function is prevalent because it converts the raw output scores (utilities) into probabilities that sum to one, making it suitable for modelling choice probabilities when assuming the error components are Independent and Identically Distributed (IID) and follow the Gumbel distribution (Wang et al., 2020a). However, this reliance on the Softmax function and its similarity to the logit formulation for probability estimation inherently limits the model to a specific distributional form for the unobserved components of utilities. Moreover, the utilities in these DNN models are commonly assumed to be independent. Those studies that have tried to introduce correlations between error components are limited to specific correlation patterns. Examples of such models include L-MNL (Sifringer et al., 2020) and ASU-DNN (Wang et al., 2020a), which also utilize the Softmax function in choice probability calculations.

This research aims to address the identified limitations in theory-driven NNs by introducing a neural network model, called RUM-NN, that is grounded in RUM theory, capable of accommodating any parametric distribution for unobserved error components, and capable of introducing correlation among error components. The proposed linear RUM-NN and its learning process are fully transparent, with all parameters retaining economic significance and behavioural interpretability. In this paper, we demonstrate RUM-NN's capability to cover different error distributions. Using synthetic datasets, we evaluate how accurately RUM-NN can recover the true patterns of the data when the error distribution follow Gumbel or Normal distributions, matching the performances of MNL and MNP. Additionally, we incorporate Cholesky decomposition (Bhat and Srinivasan, 2005) into the RUM-NN specification to capture potential correlations among utility functions. RUM-NN's flexibility extends beyond common specifications in discrete choice modelling. Leveraging DNN model features, RUM-NN can capture non-linearity between dependent and independent variables in the utility by adding fully connected layers in its architecture while maintaining its consistency with RUM. To demonstrate the superiority of RUM-NN, we conduct a comprehensive assessment using real world datasets including the Swissmetro dataset (Bierlaire et al., 2001) and the London Passenger Mode Choice dataset (LPMC) (Hillel et al., 2018). We evaluate the performance and robustness of RUM-NN across both linear and non-linear specifications under different assumptions of error distributions and compare it to their corresponding econometrics models and a conventional DNN model.

The rest of the paper is organised as follows. Section 2 provides a comprehensive literature review on modelling travel decisions using both neural network models and discrete choice models. Section 3 presents the modelling framework of RUM-NN and its mathematical representation. Section 4 presents the results of numerical experiments conducted using synthetic and real datasets. Finally, section 5 offers a summary of the study's conclusions and highlights.

## 2. Background

### 2.1. Theory-driven Models

Choice modelling has ample applications in the transport discipline (Perez-Lopez et al., 2022). Most of the proposed choice models in this field are supported by RUM (Ben-Akiva et al., 2015). According to RUM, when a decision maker faces a finite set of alternatives among which one should be selected, a rational decision maker will select the alternative with the highest utility (Ben-Akiva et al., 2015, Train, 2009). The utility received from each alternative is decomposed into (1) a deterministic observable part and (2) a stochastic unobservable part. Assume a decision-maker receives utility $U_j$ from alternative $j$ out of J available alternatives ($j \in \{1,2,...,J\}$). RUM implies a utility vector of $(U_1, ..., U_J)$ where alternative $j$ is chosen if and only if (McFadden, 1981):

$$U_j \geq U_t \quad \forall t \in \{1, 2, \ldots, J\}, j \neq t \tag{1}$$

$$U_j = V_j + \varepsilon_j \tag{2}$$

In this equation, $V_j$ is the deterministic component and $\varepsilon_j$ represents the stochastic component capturing the behavioural uncertainty and omitted variables' effect. Choice probabilities are dependent on the distribution of $\varepsilon_j$. The probability of selecting alternative $j$ can be expressed as:

$$\begin{aligned} P_j &= Prob(V_j + \varepsilon_j > V_t + \varepsilon_t, \forall t \in \{1, 2, \ldots, J\}, j \neq t) = \\ &Prob(V_j - V_t > \varepsilon_t - \varepsilon_j, \forall t \in \{1, 2, \ldots, J\}, j \neq t) = \\ &\int_\varepsilon I(V_j - V_t > \varepsilon_t - \varepsilon_j, \forall t \in \{1, 2, \ldots, J\}, j \neq t) f(\varepsilon) d(\varepsilon) \end{aligned} \tag{3}$$

In this expression, $f(\varepsilon)$ is the probability density function of $\varepsilon$. $I$ is the indicator function, that equals 1 when the expression in parentheses is true and 0 otherwise.

The distributional assumptions of $\varepsilon$ determines the model specification. If $\varepsilon$ is assumed to be IID following the Gumbel distribution (Train, 2009) the formulation collapses to the MNL model where the choice probabilities $P_j$ are calculated from equation (4) (McFadden, 1974).

$$P_j = \frac{e^{V_j}}{\sum_t e^{V_t}} \tag{4}$$

This assumption simplifies estimation but can be restrictive and unrealistic. To address this limitation, researchers have explored various non-conventional error distributions to better capture complex decision-making behaviours (Paleti, 2019). Previous research has highlighted the advantages of adopting alternative distributional assumptions including q-generalized reverse Gumbel (Chikaraishi and Nakayama, 2016), heteroskedastic extreme value (Bhat, 1995), negative exponential (Alptekinoğlu and Semple, 2016), generalized exponential (Fosgerau and Bierlaire, 2009), and negative Weibull (Castillo et al., 2008) kernel error distributions. Additional approaches involve additive combinations of Gumbel and exponential error terms (Del Castillo, 2016) and a class of asymmetric distributions (Brathwaite and Walker, 2018).

For non-conventional error distributions choice probabilities will not have a closed form formulation which necessitates simulation. One example of choice probability simulator is the smoothed AR simulator proposed by Ben-Akiva and Bolduc (1996). This method involves drawing random numbers from error terms respective distributions, followed by evaluation of utility functions. As it is shown in equation (5), simulated utilities are fed into the smoothed logit function $S$:

$$S_j = \frac{e^{U_j/\lambda}}{\sum_t e^{U_t/\lambda}} \tag{5}$$

Where $\lambda$ is a positive number specified by the modeller and is known as the scale parameter. As $\lambda$ approaches zero, matrix $S$ approaches the indicator $I$ in equation (3). To calculate the simulated probability, the whole process from generating random numbers to equation (5) is repeated multiple times, and the average value of the function $S_j$ serves as an approximation for $P_j$. The smoothed AR simulator can be applied to any choice model by simulating the utilities under any distributional assumption. However, the smoothed AR simulator introduces additional complexity to the modelling process and finding the optimum value for the parameters can become challenging.

## 2.2. Data-driven Models

In recent years, artificial intelligence tools and ML algorithms such as support vector machine (SVM), Bayesian network (BN), decision tree (DT), and random forest (RF) have become popular in choice modelling (Lu et al., 2021). ML algorithms are nonparametric models that are generally divided into supervised and unsupervised models. For choice modelling, the supervised ML methods are more applicable where available alternatives are seen as output classes. The key advantage of ML methods is their superior goodness-of-fit and prediction accuracy. For example, Diallo et al. (2022) compared DT with the Multi-Nominal Logistic Regression (MNLR) and showed that DT outperforms MNLR in prediction accuracy. Other studies have shown RF and DNN, as two widely used ML methods, have superior performance compared to econometrics models in various contexts (Hagenauer and Helbich, 2017, Lhéritier et al., 2019, van Cranenburgh et al., 2022, Wang and Ross, 2018, Zhao et al., 2020). The main disadvantages of ML models are the lack of a supporting theory leading to a lack of behavioural interpretability, and model complexity often leading to non-interpretability and overfitting issues. Hence, while ML methods have the potential to improve model accuracy compared with theory-driven econometrics approaches, they are commonly criticised for their shortcomings in explainability and interpretability.

Neural networks are widely adopted within the realm of ML due to their flexible modelling structure and the ability to capture complex nonlinear relationships within data. Neural networks make great overtures in solving a majority of problems (LeCun et al., 2015). A typical neural network consists of three components: an input layer, one or more hidden layers, and an output layer responsible for delivering classification results. The hidden layers can improve the learning ability of the model through pattern recognition and error correction. Each neuron within a neural network comprises basic processing units, represented as $f$, that transform inputs through nonlinear functions. During the training process, the model's parameters, often denoted as $\beta$, are estimated using the backward propagation learning process (Hecht-Nielsen, 1992). This estimation typically revolves around optimising an objective function, such as maximising the likelihood function. The neural network architecture can be represented as a series of chain functions as shown in equation (6).

$$z^{(1)} = f^1(V)$$

$$z^{(2)} = f^2(z^{(1)})$$

$$\ldots$$

$$P = f^l(z^{l-1})$$

(6)

In this equation, V is the input, $f^1$, $f^2$, …, $f^l$ are activation functions, $l$ represents the layers' number and $P$ is the output probabilities of the model. In discrete choice modelling, the Softmax activation function is usually used as the activation function in the last layer, as the mathematical formulation of the Softmax activation function is identical to the choice probability formulation in MNL shown in equation (4) (Qi et al., 2017).

Applications of neural networks for discrete choice modelling in the context of travel behaviour can be classified into two categories with respect to their relationship with econometric methods. The first category includes studies that use neural networks mainly as a "black box" tool to obtain superior prediction accuracy with minimal discussions around the connection to behavioural theories. For example, Assi et al. (2018) compared a DNN model with a logit model for the mode choice behaviour of high school students and concluded a higher prediction accuracy for the DNN model.

The second category includes recent studies devoted to establish connections between neural networks and statistical theories such as RUM. Wang et al. (2020a) proposed a specific neural network architecture called ASU-DNN that resembles the behaviour of MNL. ASU-DNN is a stack of fully connected subnetworks, each of which represents a utility function. The choice probabilities are then calculated using the 'Softmax' activation function in the output layer, mirroring the formulation of the probability function in MNL. The authors concluded that the results of ASU-DNN show an improvement in classification accuracy and provide more behavioural information than fully connected DNNs. Sifringer et al. (2020) introduced the Learning Multinomial Logit (L-MNL) model which also resembles the behaviour of an MNL but introduces non-linearity in utility functions through Convolutional Neural Networks (CNN). L-MNL consists of a data-driven part and a knowledge-driven part. The data-driven part includes a fully connected network with no prior relationship. On the contrary, the theory-driven part is a CNN architecture to model utility. This design allows the knowledge-driven part to offer interpretable economic insights. The authors expanded L-MNL to a nested logit model to account for correlations in the dataset. But similar to ASU-DNN, choice probabilities are obtained by applying a softmax activation function in the output layer. Han et al. (2022) extended the structure of L-MNL to a DNN model with two modules TestNet and MNL that learn the interactions between individual characteristics and alternative-specific attributes. Moreover, Wang et al. (2020b) introduced a hyperparameter to control the contribution of the data-driven and knowledge-driven parts. They also introduced the Multitask Learning Deep Neural Network (MTLDNN) as an alternative framework for implementing the nested logit model. To enhance the MTLDNNs interpretability, they calculated elasticities and visualised the relationship between the choice probabilities and the input variables. The authors showed that MTLDNN outperforms nested logit in prediction accuracy but underperforms in log-likelihood. A recent study by Bei et al. (2023) used MTLDNN to predict travel mode choice and trip purpose at the same time. Their findings demonstrated that MTLDNN outperformed both traditional DNN and MNL models in predictive accuracy. Moreover, Wong and Farooq (2021) introduced Reslogit, which is similar to L-MNL, but they extended the utility function to accommodate nonlinear forms using a series of residual layers. Kamal and Farooq (2024) extended ResLogit to the Ordinal-ResLogit to model ordinal datasets. All these studies are established on the similar mathematical equation of the softmax activation function in neural networks and the choice probability formulation in MNL.

When the Softmax activation function is applied to the last layer of neural networks, the output probabilities are identical to the choice probabilities obtained from a logit formulation (Wang et al. (2021)). Almost all existing theory-based DNN models utilize the Softmax function to calculate choice probabilities.

Designing a neural network architecture based on this similarity limits the model to specifications where the stochastic component of utilities follows a Gumbel distribution. To overcome this limitation, we propose RUM-NN, a custom-built neural network architecture that is fully compatible with RUM and capable of handling any distribution for the stochastic components. The utility formulation of the llinear RUM-NN directly parallels conventional utility functions found in the econometrics method, and it can be extended to encompass nonlinear forms. This level of transparency and the one-to-one comparability of the linear RUM-NN empower modellers to draw behavioural conclusions from RUM-NN in the same manner as they do in econometrics models. Furthermore, we have incorporated a custom-build layer within RUM-NN to implement Cholesky decomposition enabling the model to capture correlations between stochastic components.

## 3. Methodology

This section presents the custom-built architecture of RUM-NN. The details of RUM-NN with independent stochastic components are discussed in subsection 3.1, then, the implications of Cholesky decomposition to capture potential correlations between utilities are presented in subsection 3.2. Finally, the rationale behind the training process is discussed in subsection 3.4.

### 3.1. RUM-NN

#### 3.1.1. Basic structure

RUM-NN is a custom-built neural network model designed to replicate RUM. It draws inspiration from a smoothed AR simulator to simulate choice probabilities. This model has a core module that for any given instance of random terms, calculates the utilities and identifies the alternative with the highest utility. RUM-NN contains a random number generator layer, and a sufficiently large number of this core module, which enables it to simulate choice probabilities. The most basic version of RUM-NN with a linear utility structure and without correlations between utilities, comprises six layers namely the input layer, deterministic layer, random number generator layer, stochastic layer, choice layer and probability layer:

- The input layer receives the independent variables which are assumed to influence the choice. This layer has as many nodes as the number of independent variables. Each node in this layer represents one of the independent variables, serving as the initial input to the model.

- The deterministic layer linearly transforms the input variables to the deterministic utility units. This layer contains as many nodes as the number of alternatives, and each node is connected to its corresponding independent variables nodes in the input layer. This specific connection pattern resembles a linear construct for utilities. To model non-linear utility functions, this specification can be extended to multiple layers to model non-linear transformation from the input layer to the deterministic layer.

- The random number generator layer introduces stochasticity into the RUM-NN model by generating random numbers representing unobserved utility components. Each node in this layer corresponds to an alternative, and in model simulation, it produces a random value based on its specified probability distribution. This distribution can take any parametric or non-parametric form and must be defined by the modeller before training. The modeller can also fix the parameters of these nodes to obtain the desired level of identification. This means that the modeller has the flexibility to define specific characteristics of the random number distributions used in the random number generator layer. The parameters that the modeller can fix could include the mean, variance, and other moments of the distribution, depending on the chosen parametric form. The output from this layer, along with the output from the deterministic layer, feeds into the stochastic layer. This connection pattern resembles the additive inclusion of stochasticity into utility.

- The stochastic layer resembles simulated utility values. The number of nodes in this layer is the number of alternatives times the number of replications of the core module. For each replication of the core module, the nodes in the stochastic layer represent one simulated instance of utility values.

- The choice layer identifies the alternative with the highest utility for each replication of the core module. It contains as many nodes as the stochastic layer. The nodes in the choice layer take values of approximately 0 and 1. The transition from the utility layer to the choice layer for each replication of the core unit is formulated using the smoothed logit function in equation (5). Using this operator, the alternative with the maximum utility will take a value of approximately 1 and the rest of the units will take a value of approximately 0.

- The probability layer aggregates the choice layer over the replications of the core module to approximate the choice probabilities based on the number of times each alternative is observed to maximise utility. This layer has as many layers as the number of alternatives, and it represents simulated choice probabilities.

Figure 1 shows the connections between the input, deterministic, random number generator, stochastic and choice layers in one replication of the core module.

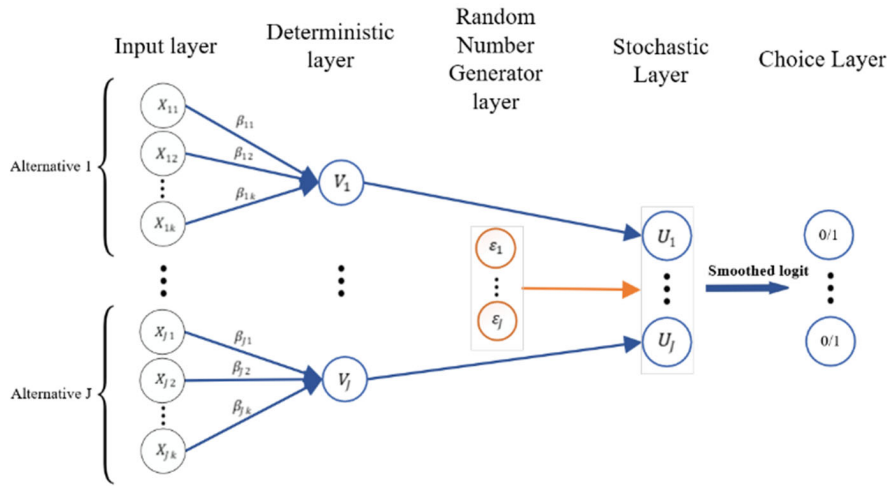

Figure 1: The connections between input layer, deterministic layer, random number generator, stochastic layer and choice layer in one replication of the core module in RUM-NN. This figure shows the model architecture up to the one-hot outputs of the choice layer. The deterministic layer takes the utility specification showing transmission of input variables into deterministic utilities. In the stochastic layer, as shown in equation (7), the error components are added to the utility. The choice layer transforms the outputs of the stochastic layer to a one-hot vector using the smoothed logit function.

### 3.1.2. Mathematical formulation

In the mathematical formulations in this section, the index for decision-makers is suppressed for ease of presentation. The stochastic components are assumed to be identical and independently distributed across decision-makers. Assume a decision-maker is faced with $J$ discrete alternatives. For each alternative, there are $K$ attributes, contributing to the observed utility associated with that alternative as $V_j = \sum_{k=1}^{K} \beta_{jk} x_{jk}$ where $\beta_{jk}$ are the parameters of the utility function and $x_{jk}$ is the input variable. To replicate the linear form of utility constructs, an identity function is used as the activation function in the deterministic layer ($f(x) = x$). The stochasticity is added to the deterministic layer in the next step. Unit values in the stochastic layer will be as shown in equation (7):

$$U_j = V_j + \varepsilon_j = \sum_{k=1}^{K} \beta_{jk} x_{jk} + \varepsilon_j \tag{7}$$

The error terms $\varepsilon_j$ can follow any parametric distribution, making the distribution of $\varepsilon_j$ a hyperparameter of RUM-NN. The distribution of $\varepsilon_j$ must be defined before training the model, and it can be specified as any parametric form, such as an Extreme Value Type I (EVI) distribution, a normal distribution, or even non-parametric distributions. The parameters of these distributions can also be specified by the modeller to achieve the desired interpretability and identification.

Figure 2 presents the steps of transmission of utilities into the probabilities. In the first step, the utility vector $[U_1, \dots, U_J]$ in the stochastic layer is transformed to a one-hot vector in the choice layer by applying the smoothed logit function $S$ in equation (5). This operator normalises values between zero and one while raising the maximum close to one and lowering the rest close to zero. This operator identifies the alternative with the highest utility. Therefore, the smoothed logit function $S$ in RUM-NN can be written as shown in equation (8).

$$S_{qj} = \frac{e^{U_{qj}/\lambda}}{\sum_t e^{U_{qt}/\lambda}}, q \in \{1, \dots, Q\} \tag{8}$$

The whole process is replicated for a sufficiently large number of times, denoted by Q (e.g., $q = 1$ shows the first repetition) to calculate the frequency of each alternative maximising utility. It should be noted that for all Q replications, $V_1, \dots, and\ V_J$ remain unchanged and only $\varepsilon_j$ from the random number generator layer vary. As mentioned in the previous section, $\lambda$ is the scale parameter. Using a sufficiently small $\lambda$, the choice layer outputs become sufficiently close to a one-hot vector indicating the alternative with maximum utility.

In the following, a stack of Q one-hot vectors is generated and stored in a $Q \times J$ matrix as shown in equation (9).

$$M = \begin{bmatrix} S_{11} & \cdots & S_{1J} \\ \cdots & \cdots & \cdots \\ S_{Q1} & \cdots & S_{QJ} \end{bmatrix}, \sum_{j=1}^{J} S_{qj} = 1, \forall q \in \{1, \ldots, Q\} \tag{9}$$

In this matrix, each column corresponds to one replication of the core module, where $S_{qj^*} \approx 1$ for $j^* = argmax_j(U_{qj})$ and $S_{qj} \approx 0$ for $\forall j \neq j^*$. Each column corresponds to an alternative, and the sum of each column reveals how many times each alternative has the maximum utility. This means the probability of alternative $j$ maximising utility can be approximated as shown in equation (10):

$$P_j(x|\beta) = \frac{\sum_{q=1}^{Q} S_{qj}}{Q} \tag{10}$$

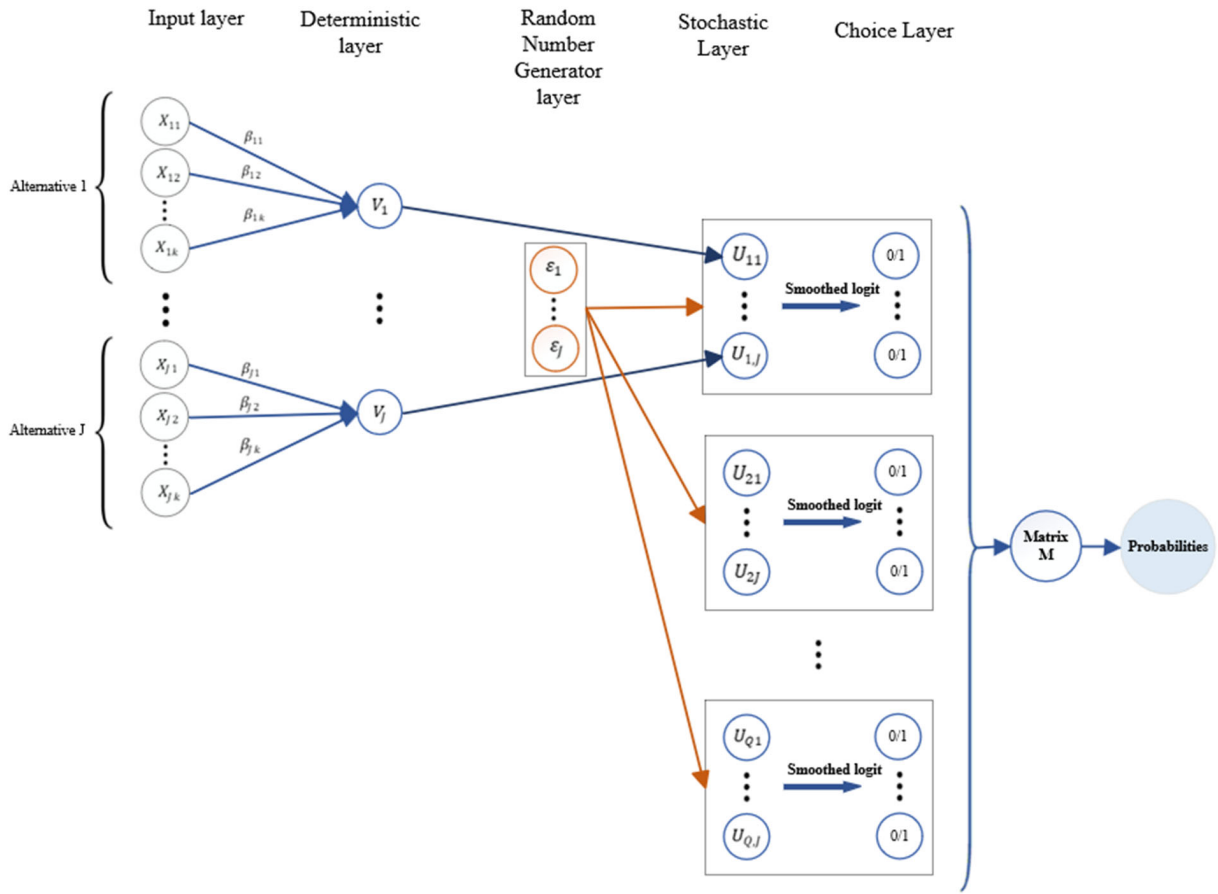

Figure 2: Overview of the probability layer and core module replications in RUM-NN. Each connection from $\varepsilon$ represents a random draw from the distribution $\varepsilon_j$ added to the deterministic layer.

RUM-NN serves as a simulator to approximate probabilities $P$ in equation (3) regardless of whether the integral has a closed-form solution. This provides full control over model specification including setting the distribution of the stochastic terms and fixing parameters for identification. Moreover, each connection in RUM-NN represents a theoretically supported relationship between variables. RUM-NN provides full explainability, enabling the calculation of any economic indicators (e.g., willingness-to-pay, and value of time) comparable to econometrics methods.

### 3.2. The correlated error terms

The behavioural RUM-NN architecture, as outlined in section 3.1, offers avenues for expansion in various directions. One such expansion entails capturing correlations among stochastic terms in utilities. Assume the error components in equation (7) are correlated across alternatives and assume $\mathbf{E}_{J \times J}$ denote their covariance matrix. It must be noted that utilities can only be measured relative to each other, and their absolute values are not identifiable. In econometrics methods, identification is typically achieved by setting one of the utilities to zero (Dansie (1985)) and other utilities are normalised accordingly. Therefore, the utility formulation of equation (7) can be rewritten as shown in equation (11). In this equation, alternative $J$ serves as the base utility.

$$U_j = V_j + \varepsilon_j^* - U_J, \qquad \varepsilon^* = \mathbf{\Psi}\,\varepsilon \tag{11}$$

Here $\varepsilon$ is assumed to be IID, and $\varepsilon^*$ is the stochastic term with covariance matrix $\mathbf{\Psi}_{(J-1)\times(J-1)}$. $\mathbf{\Psi}$ is a positive-definite covariance matrix to be estimated. The diagonal terms in $\mathbf{\Psi}$ are normalised to 1 to control the scale. The off-diagonal terms in $\mathbf{\Psi}$ represent the error covariance across alternatives. It is common to use an unconstrained parameterisation of the Cholesky matrix (Bhat and Srinivasan, 2005), in which $\mathbf{\Psi}$ is decomposed into a lower triangular matrix as shown in equation (12):

$$\mathbf{\Psi} = \mathbf{L}\mathbf{L}' \tag{12}$$

In this equation, $\mathbf{L}$ is a lower triangular matrix. As mentioned before, $\mathbf{\Psi}$ is a positive-definite matrix with diagonal values of 1. Bhat and Sidharthan (2012) proposed a parameterisation for $\mathbf{L}$ so that $\mathbf{L}\mathbf{L}'$ yields a positive-definite matrix with unit elements on its diagonal. In a scenario involving three alternatives, $\mathbf{\Psi}$ would manifest as a $2 \times 2$ matrix as shown in equation (13):

$$\mathbf{\Psi} = \begin{bmatrix} 1 & A_{12} \\ A_{12} & 1 \end{bmatrix} \tag{13}$$

The lower triangular $\mathbf{L}$ matrix is parameterised as shown in equation (14):

$$L = \begin{bmatrix} 1 & 0 \\ A_{12} & \sqrt{1 - A_{12}^2} \end{bmatrix} \qquad (14)$$

In this equation, $A_{12}$ is the parameter of the model that needs to be estimated. Using $A_{12}$, The lower triangular $L$ matrix is obtained which enables the calculation of the covariance matrix $\Psi$. This approach empowers RUM-NN to effectively capture and model correlations among the error components.

### 3.3. Non-linearity

To enhance the capability of RUM-NN in capturing complex patterns within the dataset, the structure of the deterministic layer can be expanded to include nonlinearity. This expansion involves incorporating several hidden layers within the deterministic component of utilities. The overview of the proposed nonlinear RUM-NN from the input layer to the deterministic layer is shown in Figure 3.

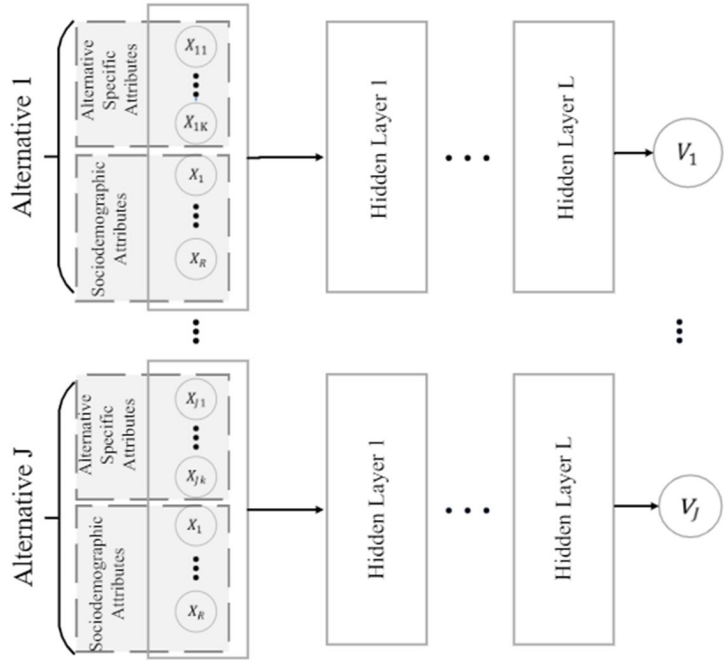

Figure 3: The structure of non-linear RUM-NN from the input variable layer to the deterministic layer

In the nonlinear version of RUM-NN, the deterministic layer no longer consists of a linear transformation. Instead, the deterministic linear utilities include several fully connected layers that create nonlinear relationships between input variables and deterministic utilities. For this purpose, the input variables are first divided into groups of alternative-specific variables $x_{jk}$ ($k \in \{1,2,\dots,K\}$ and $K$ is the number of alternative-specific variables for alternative $j$) and sociodemographic attributes $x_r$ ($r \in \{1,2,\dots,R\}$ that $R$ is the number of sociodemographic attributes). Each group, along with the sociodemographic attributes enters a separate subnetwork with fully connected hidden layers. Each hidden layer $l$ consists of $c_l$ neurons where $f^l = (f_1^l, f_2^l, \ldots, f_{c_l}^l)$ describes the relationship between parameters $w$ and the input $X$. This transformation process is represented in equation (15).

$$V_j = f^l(f^{l-1}(\ldots f^1([x_{j1}, x_{j2} \ldots, x_{jK}], [x_1, \ldots, x_R]))) \tag{15}$$

In this equation, $f^1, f^2, \ldots, f^l$ denote the activation functions used in each hidden layer, and $l$ is the number of layers. $x_{j1}, x_{j2} \ldots, x_{jK}$ represent a set of alternative specific attributes, and $x_1, \ldots, x_R$ is a set of sociodemographic attributes. The activation functions are set to Rectified Linear Unit (ReLU), defined as $f(x) = \max(0, x)$, meaning it outputs the input directly if it is positive; and zero otherwise (Goodfellow et al., 2016). The last layer only has one neuron in each subnetwork to ensure a single output corresponding to the utilities. The output of this multi-layered transformation is a set of utility values $V_1, \ldots, and\ V_J$, where each utility value is now influenced by the nonlinear interactions of the input variables. These transformed utility values are then passed on to the stochastic layer, where the error components $\varepsilon_j$ are added, as described in Section 3.1.2. The rest of the calculation for the output probabilities remains the same. The utilities, including the stochastic components, are processed through the choice layer using the smoothed logit function to identify the alternative with the highest utility. The choice probabilities are then aggregated in the probability layer, as previously outlined.

As the relationship between variables becomes nonlinear, the estimated parameters no longer correspond to single interpretable coefficients, as in traditional linear econometric models. Instead, the parameters in a nonlinear RUM-NN represent a complex mapping of inputs to outputs across multiple layers to capture intricate interactions among variables. While this modelling style may reduce direct interpretability, it can offer a more realistic model for complex datasets, as it does not impose linear interactions between variables. Besides, insights can still be obtained from these models by using post-hoc explainability techniques. Post-hoc explainability is one of the explainable AI (XAI) techniques introduced to understand, trust and manage artificial intelligence decisions (Burkart and Huber, 2021).

### 3.4. Model training and parameter estimation

In the realm of econometrics methods, the $\beta$ parameters are typically estimated by maximising the loglikelihood function $L$ as shown in equation (16):

$$\max_{\beta} L(\beta|x) = \max_{\beta} \sum_N \sum_{j=1}^{J} y_j * \log\left(P_j(x|\beta)\right) \tag{16}$$

Here $y_j$ is a binary variable that takes the value of one for the selected alternative and zero otherwise. It must be noted that parameter estimates obtained from maximising the loglikelihood function are equivalent to those obtained from minimising the negative of this function as shown in equation (17):

$$\max_{\beta} L(\beta|x) = \min_{\beta} -L(\beta|x) = \min_{\beta} -\sum_N \sum_{j=1}^{J} y_j * \log\left(P_j(x|\beta)\right) \tag{17}$$

In the reals of neural networks, the negative loglikelihood function in equation (17) is commonly referred to as cross-entropy loss function in classification tasks (Goodfellow et al., 2016). Given that RUM-NN generates simulated probabilities as its output, utilising cross-entropy for parameter estimation in this model is equivalent to employing the maximum simulated likelihood method.

We use the Adam optimisation algorithm (Diederik P. Kingma, 2015) to minimise the value of the loss function $-L(\beta|x)$. The values of $\beta$ parameters will be estimated through the backward propagation process (Hecht-Nielsen, 1992). During this process, the algorithms update values of parameters $\beta$ through several steps in the direction that minimises the value of the loss function $-L(\beta|x)$.

## 4. Experimental Results

This section demonstrates how under certain error term distributions, RUM-NN can capture Gumbel and Normal error terms in parameter recovery. Synthetic datasets with known true parameters are used to assess parameter recovery. The Montecarlo simulation is utilised to show the robustness of parameter recovery. Then the performance in prediction accuracy is measured on a real-world dataset. The codes are implemented in the Keras Python library (Chollet, 2015). The analyses in this section are run on a system with a CPU of Intel(R) Core(TM) i7-9750H, a GPU of Nvidia GForce GTX1650, and a RAM of 16 GB DDR4.

### 4.1. The synthetic dataset constructs

The synthetic datasets are generated following an approach introduced by Guevara (2015) and adapted by Sifringer et al. (2020). Guevara (2015) introduced this procedure for binary choice scenarios. We have modified this approach to create datasets with two and three alternatives. The process of generating a synthetic dataset is indicated in equations (18) to (22).

$$U_j = V_j + \varepsilon_j \tag{18}$$

$$V_j = \beta_p \cdot p_j + \beta_a \cdot a_j + \beta_b \cdot b_j + \beta_q \cdot q_j \tag{19}$$

$$p_j = 5 + z_j + 0.03wz_j + \varepsilon_{pj} \tag{20}$$

$$q_j = 2h_j + k_j + \varepsilon_{qj} \tag{21}$$

$$k_j = h_j + \varepsilon_{kj} \tag{22}$$

In these equations, $\beta_p$, $\beta_a$, $\beta_b$ and $\beta_q$ are the parameters and $a_j$, $b_j$, $z_j$, $wz_j$ and $h_j$ are input variables drawn from a random Uniform distribution (-1,1). The error components including $\varepsilon_{pj}$, $\varepsilon_{qj}$ and $\varepsilon_{kj}$ are drawn from a random Uniform distribution (-1,1). $\varepsilon_j$ is the stochastic component of the utility with a specific distribution determined based on the experiment's objective. The synthetic choice set $y_i$ is calculated as shown in equation (23).

$$y_i = argmax\ U_i \tag{23}$$

## 4.2. Simulation

We conduct three experiments each designed to test different aspects of the model's performance. The first experiment involves 100 datasets with binary choices where the error components follow an IID Gumbel distribution, designed to compare the performance of RUM-NN with Gumbel distribution error terms against the traditional logit model. The second experiment uses 100 generated datasets with binary choices where the error components follow a normal distribution. Through this experiment, the performance of RUM-NN with normally distributed error terms is compared against the probit model. The third experiment involves 100 datasets with three alternatives, where the error terms are normally distributed and correlated. This analysis illustrates the model's performance transition from a dataset with two alternatives and non-correlated error terms to a dataset with three alternatives and correlated error terms. Details of each experiment are discussed in the following subsections, and the analysis of hyperparameters is presented in Subsection 4.3.2.

### 4.2.1. Experiment I: IID Gumbel distribution

The synthetic dataset in experiment I is generated using the process of data generation outlined in Section 4.1. Specifically, the values of $\beta_p, \beta_a, \beta_b$ and $\beta_q$ are set to -1, 0.5, 0.5, and 1 respectively, and 1000 binary choice records are generated. Monte Carlo simulations are applied to generate 100 datasets. For this experiment, the error component $\varepsilon_j$ in equation (18) is assumed to follow a Gumbel distribution with location and scale parameters of zero and one, respectively.

Each of these synthetic datasets is used to train a RUM-NN, an MNL model, and an MNP model. The average values and standard deviations of the estimated parameters from the Monte Carlo experiments are presented in Table 1. The results indicate that the performance of RUM-NN in parameter recovery is comparable to that of the MNL model, as both models effectively capture the Gumbel distribution of the error components in the datasets. This similarity suggests that RUM-NN is robust in terms of parameter estimation when the underlying error structure aligns with the assumptions of the MNL model. However, the MNP model struggled to recover the true parameters accurately, reflecting a mismatch in scale due to the misspecification of assuming normally distributed error terms. These findings underscore the importance of aligning model assumptions with the underlying data characteristics, demonstrating that RUM-NN can provide accurate parameter estimates under conditions traditionally assumed for the MNL model while also offering flexibility for different error structures.

Table 1: The average estimated parameters from the Montecarlo experiment for the MNL, MNP and RUM-NN models. The unobserved error terms for data generation and RUM-NN estimation are drawn from the Gumbel distribution.

| Parameters | True value | MNL | | MNP | | RUM_NN | |
|---|---|---|---|---|---|---|---|
| | | Estimated value | Std | Estimated value | Std | Estimated value | Std |
| $\beta_p$ | -1 | -1.00 | 0.03 | -0.56 | 0.01 | -1.00 | 0.04 |
| $\beta_a$ | 0.5 | 0.50 | 0.03 | 0.28 | 0.02 | 0.50 | 0.04 |
| $\beta_b$ | 0.5 | 0.50 | 0.03 | 0.28 | 0.02 | 0.50 | 0.04 |
| $\beta_q$ | 1 | 1.00 | 0.02 | 0.56 | 0.01 | 1.00 | 0.03 |

It is noteworthy that RUM-NN offers full control over the details of utilities' stochastic terms. In DCM, the scale of the error term is fixed for identification. Almost all the existing packages to develop an MNL set the variance of the error term to $\frac{\pi^2}{6}$. In experiment I, we used the same variance when constructing the datasets, and when developing RUM-NN. This ensures that the estimated parameters from RUM-NN are directly comparable with their true values. This feature eliminates the need to compare ratios of parameter estimates, as is sometimes necessary in previous studies (e.g. (Sifringer et al., 2020)).

### 4.2.2. Experiment II: IID Normal distribution

In experiment I, the performance of RUM-NN in parameter recovery and prediction accuracy is compared against MNL. While MNL is only applicable when the error terms ε (equation (2)) are IID and follow the Gumbel distribution, RUM-NN is not limited to a specific distribution and can be applied to datasets with any parametric or non-parametric distributions. In experiment II, the performance of RUM-NN in parameter recovery and prediction accuracy is examined when the error terms ε follow the Normal distribution. In this experiment, the performance of RUM-NN is compared against MNP, which is its counterpart in the econometrics realm when the error term ε is assumed to be Normally distributed. In this experiment, 100 datasets with 10000 observations and binary outputs are generated using the procedure discussed in Section 4.1, and for all datasets, the error component is drawn from the Normal distribution with a mean of zero and standard deviation of 1. Similar to the previous experiment, $\beta_p$, $\beta_a$, $\beta_b$ and $\beta_q$ are set to -1, 0.5, 0.5, and 1 respectively.

Table 2: The average estimated parameters from the Montecarlo experiment for the MNL, MNP and RUM-NN models. The unobserved error terms for data generation and RUM-NN estimation are drawn from the Normal distribution.

| Parameters | True value | MNL | | MNP | | RUM_NN | |
|---|---|---|---|---|---|---|---|
| | | Estimated value | Std | Estimated value | Std | Estimated value | Std |
| $\beta_p$ | -1 | -1.62 | 0.05 | -1.00 | 0.02 | -1.00 | 0.03 |
| $\beta_a$ | 0.5 | 0.81 | 0.04 | 0.50 | 0.02 | 0.50 | 0.03 |
| $\beta_b$ | 0.5 | 0.81 | 0.04 | 0.49 | 0.03 | 0.49 | 0.03 |
| $\beta_q$ | 1 | 1.63 | 0.03 | 1.00 | 0.01 | 1.00 | 0.02 |

The average values and standard deviations of the estimated parameters from the Monte Carlo simulations are presented in Table 2. As shown in the table, RUM-NN successfully recovered all true parameters with performance comparable to that of the MNP model. In contrast, the MNL model was unable to recover the true parameters due to the misspecification of assuming Gumbel distribution for the error terms. This experiment highlights the flexibility of RUM-NN in handling different error term distributions, making it a viable alternative to MNP in discrete choice modelling. As discussed in Section 4.2.1, RUM-NN allows the modeller to specify the distribution and scale of the error term. In this experiment, we fixed the scale to 1 and considered the Normal distribution for the error term, ensuring that all parameter estimates are directly comparable to their true values. These findings demonstrate the robustness and adaptability of RUM-NN in modelling discrete choices under varying error structures.

### 4.2.3. Experiment III: Correlated distribution

Experiment III demonstrates the capability of RUM-NN to model discrete choice scenarios with correlated error terms. To generate synthetic datasets for this experiment, the procedure described in Section 4.2.2 is extended to include three alternatives with correlated error terms. Specifically, correlation is introduced between the error terms of the first and second alternatives. We use the Cholesky decomposition method discussed in Section 3.2 to capture correlation among the error terms, as obtained in equation (11).

For identification purposes, the error term in the third alternative is set to zero, and correlation is introduced between the error terms of the first and second alternatives. In this experiment, $A_{12}$ (equation (13)) is assumed to be 0.4, and $\varepsilon_j$ are drawn from the Normal distribution with a mean of zero and a variance of one. Similar to the previous experiments, 100 datasets with 10,000 observations are generated, and the performance of RUM-NN in parameter recovery is compared with that of MNP, and MNL.

The mean and standard deviation of the estimated parameters for MNL, MNP, and RUM-NN are reported in Table 3. According to the results, RUM-NN's performance in parameter recovery is comparable to that of MNP. As shown in the table, the MNL model could not accurately recover the true parameters and failed to account for the correlation between error components. This experiment underscores RUM-NN's ability to capture correlation among error terms. Notably, RUM-NN successfully recovered the correlation parameter, $A_{12}$, highlighting its unique capability in this regard. This distinctive feature sets RUM-NN apart from previous attempts to apply utility maximisation theory to neural networks, demonstrating its potential as a robust and flexible tool in discrete choice modelling.

Table 3: The average estimated parameters from the Montecarlo experiment for the MNL, MNP and RUM-NN models. The unobserved error terms are correlated and normally distributed.

| Parameters | True value | MNL | | MNP | | RUM_NN | |
|---|---|---|---|---|---|---|---|
| | | Estimated value | Std | Estimated value | Std | Estimated value | Std |
| $\beta_p$ | -1 | -1.79 | 0.03 | -1.00 | 0.03 | -1.00 | 0.03 |
| $\beta_a$ | 0.5 | 0.89 | 0.03 | 0.50 | 0.02 | 0.50 | 0.02 |
| $\beta_b$ | 0.5 | 0.89 | 0.04 | 0.50 | 0.02 | 0.50 | 0.02 |
| $\beta_q$ | 1 | 1.80 | 0.02 | 1.00 | 0.02 | 1.00 | 0.02 |
| Correlation alternative1-alternative2 | 0.4 | - | - | 0.39 | 0.03 | 0.38 | 0.04 |

### 4.3. Empirical application

In this subsection, the performance of RUM-NN and MNL on the Swissmetro dataset (Bierlaire et al., 2001) and LPMC (Hillel et al., 2018) is evaluated. An overview of Swissmetro dataset is presented in Section 4.3.1, and the results are detailed in Section 4.3.1.1 and 4.3.1.2. the LPMC dataset is described in Section 4.3.2, and the results are reported in Section 4.3.2.1 and 4.3.2.2. The hyperparameters used in this analysis are thoroughly examined in Section 4.3.2, providing insights into their impact on model performance.

### 4.3.1. Swissmetro dataset

The swissmetro dataset was collected from passengers between St. Gallen and Geneva, Switzerland in March 1998 (Bierlaire et al., 2001). Respondents were surveyed about their preferred travel mode. The dataset contains 10,728 observations in total. After cleaning the dataset, 9,036 observations are included in the modelling dataset of this study. This dataset is openly available, ensuring that our research can be reproduced and validated by other researchers.

#### 4.3.1.1. Model Performance

To assess goodness-of-fit, we use log-likelihood and prediction accuracy. The error term distributions in RUM-NN are set to Gumbel, Normal, Exponential, and Pareto for both linear and nonlinear structures. Gumbel and Normal distributions are well-known and widely used in discrete choice modelling. In addition to these conventional distributions, we decided to include a non-conventional distribution, Exponential, which has been used in prior studies e.g. (Fosgerau and Bierlaire, 2009, Alptekinoğlu and Semple, 2016, Del Castillo, 2016). Furthermore, we introduced the Pareto distribution, which, to the best of our knowledge, has not been previously explored in choice modelling for the error term. The probability density functions (PDFs) of the Exponential and Pareto distributions are presented in equations (24) and (25), respectively.

$$\varepsilon_{Exponential}(x, \gamma) = \begin{cases} \gamma e^{-\gamma x}, & x \geq 0 \\ 0, & x < 0 \end{cases} \qquad (24)$$

$$\varepsilon_{Pareto}(x, C, \delta) = \begin{cases} \dfrac{\delta C^\delta}{x^{\delta+1}}, & x \geq C \\ 0, & x < C \end{cases} \qquad (25)$$

Here, $\gamma$ represents the rate parameter of $\varepsilon_{Exponential}$, which is assumed to be 1. $\delta$ denotes the concentration parameter, and $C$ is the scale parameter, both of which are also assumed to be 1.

We perform a 5-fold cross-validation to ensure robustness in our results. In 5-fold cross-validation, the model is trained and tested five times, each time using a different part as the test set and the remaining parts as the training set. Subsequently, RUM-NN, DNN, MNL and MNP models are calibrated on the training dataset, and the in-sample and out-sample log-likelihood and prediction accuracy are calculated and reported in Table 4. The average of 5-fold results is reported to provide a more reliable estimate of the model's performance.

The nonlinear models (non-linear RUM-NN and DNN) consistently outperform linear models (linear RUM-NN, MNL and MNP), as shown by their higher log-likelihood and prediction accuracy. This observation demonstrates the ability of nonlinear models to capture the complex relationships present in real-world datasets, which often exhibit intricate interactions and nonlinear dependencies.

Among the linear models, RUM-NN with Gumbel and MNL, and RUM-NN with Normal and MNP, exhibit similar performance, as expected. This result is intuitive, as linear RUM-NN replicates the structures of MNL and MNP when paired with the Gumbel and Normal distributions, respectively. Although the differences in performance among the linear models are marginal, the performance of RUM-NN with Gumbel and Exponential distributions is slightly superior to that with Normal and Pareto distributions. For the nonlinear models, RUM-NN with all four distributions achieves superior test and train accuracy compared to the standard DNN model. Notably, RUM-NN with the Pareto distribution achieves the highest training and testing accuracy, at 77.41% and 72.16%, respectively.

These findings emphasise the importance of considering both the utility structure and the error distribution when designing discrete choice models. The nonlinear RUM-NN models outperform not only because of their ability to model complexity but also because they integrate error distributions effectively. The results also encourage further exploration of non-standard error distributions, particularly in applications where datasets exhibit unique characteristics that traditional distributions may fail to capture.

Table 4: In-sample and out-sample goodness-of-fit measures of the MNL, MNP, DNN and RUM-NN models for the Swissmetro dataset.

| Specification | Utility Structure | Error Distribution | $Log-likelihood$ | | $Accuracy$ | |
|---|---|---|---|---|---|---|
| | | | $Train$ | $Test$ | $Train$ | $Test$ |
| $MNL$ | Linear | Gumbel | -4603.07 | -1155.82 | 66.31 | 66.21 |
| $MLP$ | Linear | Normal | -4651.55 | -116940 | 66.08 | 65.79 |
| $RUM\_NN$ | Linear | Gumbel | -4610.41 | -1155.04 | 66.94 | 66.81 |
| $RUM\_NN$ | Linear | Normal | -4656.92 | -1173.17 | 65.42 | 65.33 |
| $RUM\_NN$ | Linear | Pareto | -4649.97 | -1166.96 | 65.59 | 65.48 |
| $RUM\_NN$ | Linear | Exponential | -4614.34 | -1157.04 | 66.65 | 66.57 |
| $DNN$ | Non-Linear | Gumbel | -3460.52 | -1148.41 | 74.15 | 70.73 |
| $RUM\_NN$ | Non-Linear | Gumbel | -3435.95 | -1047.45 | 75.31 | 71.66 |
| $RUM\_NN$ | Non-Linear | Normal | -3440.60 | -1111.68 | 75.05 | 70.56 |
| **$RUM\_NN$** | **Non-Linear** | **Pareto** | **-3439.55** | **-993.74** | **77.41** | **72.16** |
| $RUM\_NN$ | Non-Linear | Exponential | -3398.66 | -1032.54 | 76.72 | 71.72 |

### 4.3.1.2. Validating Parameter Estimation

In this subsection, we aim to validate the alignment of parameter estimation between the linear RUM-NN with a Gumbel error distribution and the equivalent econometric model, MNL. The estimated parameters of the linear RUM-NN and MNL for the Swissmetro dataset are reported in Table 5. This table provides parameter estimates obtained through MNL and linear RUM-NN with Gumbel error term distribution. Additionally, it includes the ratio of differences between the estimated values. It is important to note that the true parameters of this dataset are unknown, so there is no ground truth to evaluate the parameter recovery performance directly. However, we can compare the parameter estimates across the models to investigate if RUM-NN with IID Gumbel error terms can yield the same results as MNL. To this end, we conduct both the Two One-Sided Test for Equivalence (TOST) and a T-test between the estimated parameters of the linear RUM-NN with the Gumbel distribution and the MNL model. These tests are aimed at determining if there are any meaningful differences between the two models. The TOST, as proposed by Seaman and Serlin (1998), involves performing two separate one-sided tests using the mean differences. The detailed results of this analysis are presented in Appendix I.

For the T-test, the hypothesis is that there are no significant differences between the estimated values of linear RUM-NN and MNL. According to the results, as all the t-values are higher than the significance level of 0.05, we accept the hypothesis that the performance of these two models is similar. In the TOST analysis, the null hypothesis is that the two models are not equivalent. The results show that these combined p-values were less than the significance level of 0.05 for all parameters. Consequently, we rejected the null hypothesis of non-equivalence and concluded that the linear RUM-NN model with the IID Gumbel distribution is statistically equivalent to the MNL model.

Table 5: parameter estimation of MNL and RMU-NN with Gumbel distributions on Swissmetro Dataset

| Parameters / Distributions | RUM-NN Gumbel | | MNL Gumbel | |
|---|---|---|---|---|
| | Average | STD | Average | STD |
| $ASC\_SM$ | -1.473 | 0.126 | -1.544 | 0.161 |
| $ASC\_Train$ | -2.373 | 0.116 | -2.399 | 0.082 |
| $\beta_{Cost}$ | -0.007 | 0.000 | -0.007 | 0.000 |
| $\beta_{Time}$ | -0.012 | 0.001 | -0.012 | 0.000 |
| $\beta_{Age\_Train2}$ | 0.516 | 0.083 | 0.531 | 0.072 |
| $\beta_{Age\_SM2}$ | 0.799 | 0.115 | 0.753 | 0.182 |
| $\beta_{Age\_Train3}$ | 0.689 | 0.121 | 0.685 | 0.087 |
| $\beta_{Age\_SM3}$ | 0.669 | 0.116 | 0.575 | 0.177 |
| $\beta_{Age\_Train4}$ | 0.996 | 0.096 | 1.038 | 0.091 |
| $\beta_{Age\_SM4}$ | 0.365 | 0.076 | 0.363 | 0.184 |
| $\beta_{Age\_Train5}$ | 1.011 | 0.086 | 1.049 | 0.141 |
| $\beta_{Age\_SM5}$ | -0.145 | 0.057 | -0.122 | 0.086 |
| $\beta_{Income\_Train2}$ | 0.245 | 0.063 | 0.202 | 0.087 |
| $\beta_{Income\_SM2}$ | 0.890 | 0.056 | 0.889 | 0.058 |
| $\beta_{Income\_Train3}$ | -0.074 | 0.032 | -0.137 | 0.136 |
| $\beta_{Income\_SM3}$ | 1.060 | 0.084 | 1.056 | 0.086 |

### 4.3.2. The London Passenger Mode Choice dataset

The LPMC dataset is based on survey data and includes features of travel modes of walking, cycling, driving, and public transport, along with passenger characteristics (Hillel et al., 2018). Alternative-specific variables, such as travel time and cost for each mode, were obtained using the Google application programming interface (API). Passenger characteristics, including age and gender, were collected through survey responses. The dataset contains 81,086 observations which is much bigger than Swissmetro dataset. A detailed description of which is provided by Hillel et al. (2018).

#### 4.3.2.1. Model Performance

This section evaluates the performance of the proposed RUM-NN model using the LPMC dataset and compares it against MNL, MNP, and DNN models. The evaluation metrics include log-likelihood and prediction accuracy, calculated through a 5-fold cross-validation process.

Table 6 summarises the log-likelihoods and prediction accuracies for the RUM-NN, DNN, MNL, and MNP models on both test and train datasets. The results demonstrate that nonlinear models generally outperform linear models in terms of both metrics, highlighting their ability to capture the complexity inherent in real-world datasets.

For linear modelling, the performance of RUM-NN with Gumbel and Normal error distributions closely aligns with that of the MNL and MNP models, respectively. This similarity is expected, as the linear RUM-NN replicates the structures of MNL and MNP when configured with these error distributions. For nonlinear modelling, RUM-NN is evaluated with various error distributions and compared to the conventional DNN model. The results indicate that RUM-NN with a Gumbel distribution outperforms the DNN model in both prediction accuracy and log-likelihood. Among the tested distributions, RUM-NN with an Exponential distribution achieves the highest accuracy, further emphasising the importance of exploring alternative error structures. Specifically, the train accuracy of RUM-NN with an Exponential distribution reaches 67.92%, surpassing the DNN model's 65.79%. Similarly, the test accuracy of RUM-NN with an Exponential distribution is 67.54%, outperforming not only the DNN model but also the MNL and MNP models.

The results from the LPMC dataset reinforce the findings from the Swissmetro benchmark, highlighting the superiority of the RUM-NN models in terms of prediction accuracy and loglikelihood. Moreover, the flexibility of RUM-NN to assume various error distributions allows it to outperform traditional models and conventional DNNs, particularly when optimised with suitable distributional assumptions.

Table 6: In-sample and out-sample goodness-of-fit measures of the MNL, MNP, DNN and RUM-NN models for the LPMC dataset.

| Specification | Utility Structure | Error Distribution | $Log-likelihood$ | | $Accuracy$ | |
|---|---|---|---|---|---|---|
| | | | Train | Test | Train | Test |
| MNL | Linear | Gumbel | -53775.70 | -13449.10 | 64.32 | 64.22 |
| MLP | Linear | Normal | -54452.26 | -13631.28 | 64.15 | 64.07 |
| RUM_NN | Linear | Gumbel | -53765.20 | -13448.70 | 64.35 | 64.34 |
| RUM_NN | Linear | Normal | -54434.27 | -13614.34 | 64.85 | 64.84 |
| RUM_NN | Linear | Pareto | -54447.12 | -13618.56 | 64.76 | 64.73 |
| RUM_NN | Linear | Exponential | -53987.70 | -13497.90 | 64.96 | 64.90 |
| DNN | Non-Linear | Gumbel | -51870.10 | -13023.80 | 65.79 | 65.63 |
| RUM_NN | Non-Linear | Gumbel | -49837.80 | 12600.10 | 67.79 | 67.39 |
| RUM_NN | Non-Linear | Normal | -49736.01 | -12587.36 | 67.80 | 67.45 |
| RUM_NN | Non-Linear | Pareto | -50520.57 | -12739.59 | 67.69 | 67.31 |
| **RUM_NN** | **Non-Linear** | **Exponential** | **-49680.23** | **-12596.41** | **67.92** | **67.54** |

### 4.3.2.2. Validating Parameter Estimation

This subsection demonstrates the alignment between the linear RUM-NN with a Gumbel distribution and the MNL model. Table 7 presents the estimated parameters for linear RUM-NN and MNL models. The purpose of this comparison is to validate the ability of RUM-NN to replicate the parameter structure of the MNL model, using real-world datasets. For the Gumbel distribution, the linear RUM-NN closely replicates the parameter estimates obtained by the MNL model. This alignment is as expected, given that RUM-NN with a Gumbel distribution theoretically replicates the MNL framework. The similarity between the parameter estimates is confirmed through statistical analyses, including the T-test and TOST. These analyses, detailed in Appendix I, demonstrate that the estimated parameters from RUM-NN and MNL are statistically equivalent, providing further validation of RUM-NN's capability to replicate MNL results.

Table 7 highlights the parameter estimates in MNL and linear RUM-NN. As shown in this table the differences between the models are minimal. This consistency shows the robustness of RUM-NN in approximating MNL results, ensuring its applicability as a flexible yet theoretically grounded modelling approach.

Table 7: parameter estimation of MNL and RMU-NN with Gumbel distribution on LPMC Dataset

| Parameters / Distributions | RUM-NN Gumbel | | MNL Gumbel | |
|---|---|---|---|---|
| | Average | STD | Average | STD |
| $ASC\_CYCLING$ | -4.547 | 0.029 | -4.506 | 0.039 |
| $ASC\_PT$ | -2.706 | 0.019 | -2.677 | 0.022 |
| $ASC\_DRIVING$ | -2.036 | 0.028 | -2.040 | 0.024 |
| $\beta_{Time\_Walking}$ | -8.304 | 0.026 | -8.300 | 0.036 |
| $\beta_{Time\_Cycling}$ | -5.162 | 0.022 | -5.142 | 0.016 |
| $\beta_{Time\_Driving}$ | -4.23 | 0.023 | -4.246 | 0.018 |
| $\beta_{Cost\_Driving}$ | -0.100 | 0.005 | -0.101 | 0.002 |
| $\beta_{Cost\_PT}$ | -0.187 | 0.008 | -0.194 | 0.003 |
| $\beta_{Time\_PT\_BUS}$ | -2.025 | 0.022 | -1.995 | 0.024 |
| $\beta_{Time\_PT\_Rail}$ | -1.632 | 0.045 | -1.617 | 0.047 |
| $\beta_{Time\_PT\_Access}$ | -4.842 | 0.018 | -4.863 | 0.026 |
| $\beta_{Time\_PT\_INT\_Wait}$ | -4.342 | 0.101 | -4.297 | 0.095 |
| $\beta_{Traffic\_Driving}$ | -3.022 | 0.032 | -2.995 | 0.035 |
| $\beta_{Age\_Driving}$ | 0.008 | 0.001 | 0.008 | 0.000 |
| $\beta_{Age\_PT}$ | -0.006 | 0.001 | -0.006 | 0.000 |
| $\beta_{Age\_Cycling}$ | -0.001 | 0.002 | -0.001 | 0.000 |
| $\beta_{Gender\_Driving}$ | -0.193 | 0.010 | -0.186 | 0.005 |
| $\beta_{Gender\_PT}$ | -0.212 | 0.018 | -0.194 | 0.003 |
| $\beta_{Gender\_Cycling}$ | -1.167 | 0.025 | -1.172 | 0.024 |

### 4.4. Hyperparameter analysis

In this section, we conduct a thorough analysis of the sensitivity of key hyperparameters within the RUM-NN model. These hyperparameters are important for optimising the model's performance and can be divided into two main categories. The first category pertains to the architectural elements of RUM-NN, including the distribution of error terms, the smoothed logit function parameter ($\lambda$) and the number of replications of the core module in the model ($Q$). The second category involves hyperparameters related to the training process, such as the learning rate, number of hidden layers and number of units in each hidden layer. The impact of error component distributions is comprehensively discussed in Section 4.2. This section focuses on evaluating how variations in $\lambda$ and $Q$ affect the model's performance. By systematically examining these hyperparameters, we aim to provide insights into their roles and optimise their settings to enhance the robustness and accuracy of the RUM-NN model.

When calculating choice probabilities in RUM-NN, two approximations come into play. First, the application of the smoothed logit function transfers utilities into a one-hot vector, and second, the core module is replicated to approximate choice probabilities. The hyperparameters of $\lambda$ and $Q$ give the modeller control over these approximations. Finding appropriate values for $\lambda$ and $Q$ is crucial for RUM-NN performance. If $\lambda$ is set too high, the one-hot vector may not serve its purpose in identifying the alternative with maximum utility, leading to inaccurate probabilities. Conversely, if $\lambda$ is set too low (very close to zero), numerical limitations may cause exact zero and one values in the one-hot vector, disrupting the parameter estimation process. This condition is not desirable as the smoothed logit function will collapse to the argmax operator for which the gradient is not defined. This disrupts the parameter estimation process. As discussed before, the backpropagation method used to estimate the parameters evaluates the expression for derivatives of the loss function as a product of derivatives between each layer. Therefore, finding an appropriate value for $\lambda$ is essential for a desirable performance from RUM-NN. In the experiments in Section 4.2, we set $\lambda = 1e - 4$.

The second hyperparameter, $Q$, controls the number of replications. To illustrate the impact of $Q$ on parameter estimation, the Monte Carlo simulation of experiment I is repeated for various values of $Q \in \{10, 100, 500, 1000\}$. Figure 4 shows the distribution of the estimated parameters for different values of $Q$. Each plot in Figure 4 corresponds to one of the four model parameters, with the true values indicated by a horizontal dashed line, and the shaded area representing one standard deviation around the mean estimated value. According to these plots, the mean parameter estimate is sufficiently close to the true value for Q larger than 500. This observation is limited to the dataset and model specification of experiment I. Also, as expected, the range of estimated values is wider for lower values of Q, and the range becomes narrower as Q increases.

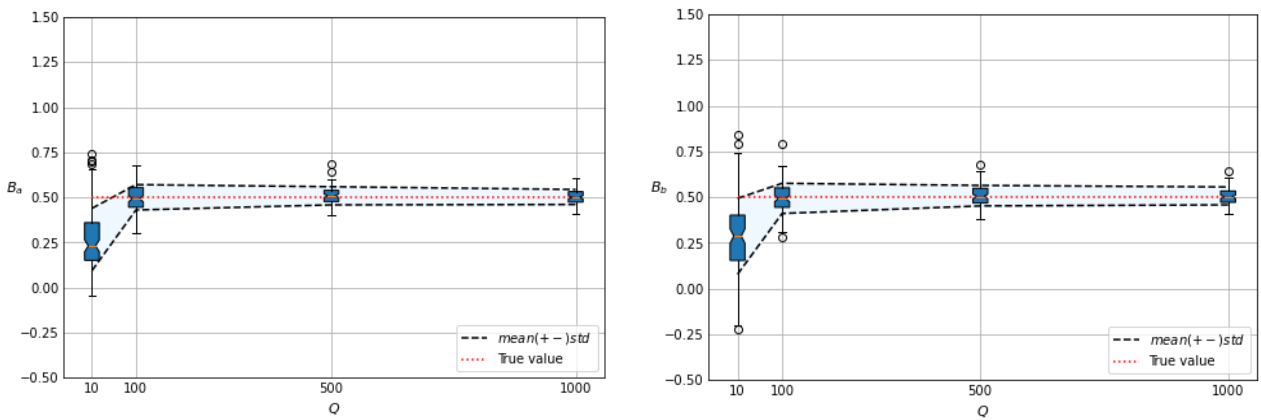

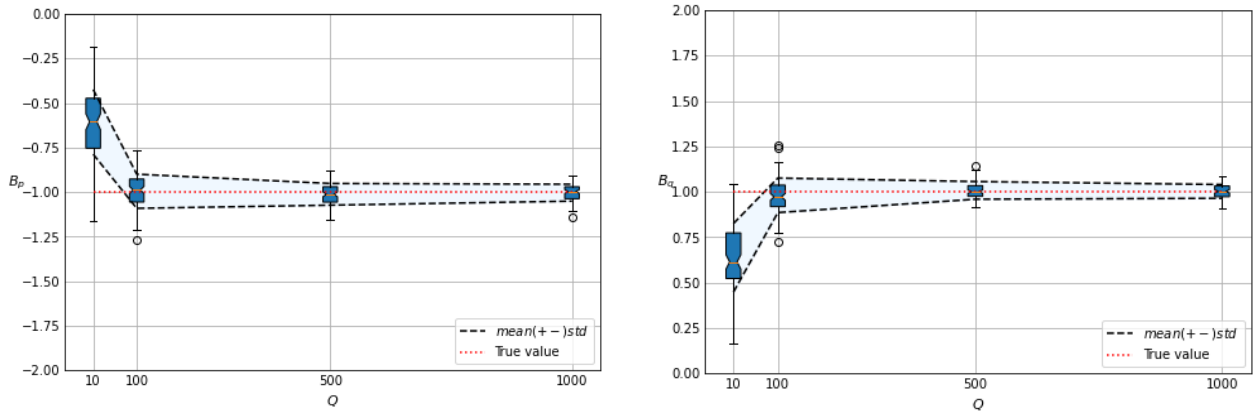

Figure 4: Boxplots of estimating parameters $\beta$ under the different values of $q$

The plots in Figure 4 do not suggest a considerably large difference in parameter recovery for Q higher than 500; however, the appropriate value of $Q$ depends on the complexity of the dataset. On the other hand, selecting an unnecessarily large Q can impose high computational costs. To investigate this effect, the computational time to develop a RUM-NN model with different Q values varying as $Q \in \{100, 1000, 3000, 5000, 7000, \text{and } 10000\}$ is recorded for a randomly selected dataset from experiment I. Figure 5 shows the variations in computational time over the examined range of $Q$. As shown in Figure 5, there is a linear relationship between the value of $Q$ and the computational time. The computational time significantly increases from 1000 to 10000, while according to Figure 5, the parameter recovery performance does not change noticeably after Q>500.

RUM-NN calibration can require relatively high computational time due to the numerical approximation for choice probabilities. For specifications with closed-form formulations, the calibration run time in RUM-NN will be longer than the maximum likelihood estimation (MLE), however for more complex specifications where numerical approximations are inevitable in MLE, the run times are expected to be closer.

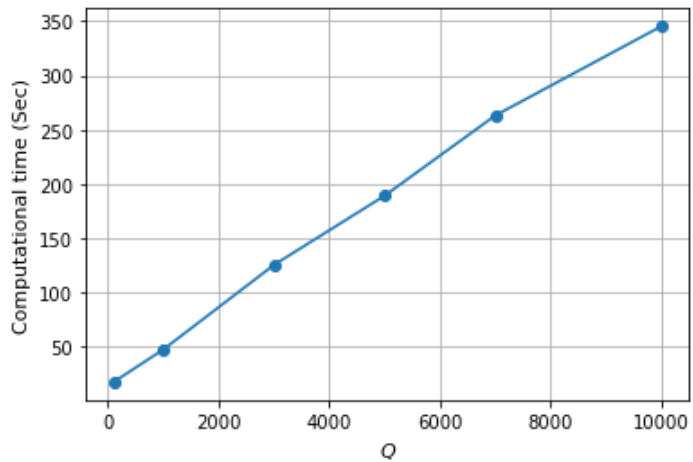

Figure 5: Computational time according to the different values of $Q$ in a stable condition

The second group of hyperparameters includes the learning rate, the number of epochs, also number of hidden layers, and the number of units in the case of non-linear RUM-NN. The learning rate and the number of epochs hyperparameters have been analysed in previous publications (e.g., Smith, 2018). In this study, we set the learning rate and number of epochs to 0.001 and 1000, respectively. As RUM-NN has a sparse architecture and offers a systematic mechanism to ensure identifiability, hyperparameters such as regularisation steps, commonly used to control overfitting, are not considered in this study.

To evaluate the optimal number of hidden layers in the nonlinear RUM-NN model, we analysed this hyperparameter using prediction accuracy as the performance metric. For this purpose, we considered models with 1, 2, 3, 4, 6, and 10 hidden layers. Using the Swissmetro and LPMC datasets, we employed a nonlinear RUM-NN with a Gumbel distribution to determine the most suitable number of hidden layers based on prediction accuracy and model complexity. To balance model complexity and prediction accuracy, we chose a structure with two hidden layers and 100 neurons in each layer. This configuration provides optimal performance while avoiding the increased complexity and risk of overfitting associated with adding more layers or neurons.

## 5. Conclusion

This research introduced RUM-NN, an interpretable and identifiable structure of neural networks based on RUM theory. The custom-built architecture of RUM-NN mirrors the utility construct in RUM. The structure of RUM-NN draws inspiration from the smoothed AR simulator. This design grants full control over model identifiability and empowers the calculation of economic measures akin to traditional econometrics methods. RUM-NN is designed to numerically approximate choice probabilities, incorporating a random number generator layer that can generate random numbers following any desired distribution. Therefore, similar to the smoothed AR simulator, the stochastic component of RUM-NN is highly adaptable and capable of modelling a wide range of distributions. Moreover, RUM-NN can capture correlations among error components through the implementation of the Cholesky decomposition in its stochastic layer. The proposed model, RUM-NN, can be designed in both linear and nonlinear structures. The results show that RUM-NN not only replicates the performance of MNL and MNP models but also achieves the higher prediction accuracy and log-likelihood by assuming different distributions for the error term. This flexibility and robustness make RUM-NN a powerful tool for discrete choice modelling, capable of handling complex data structures and providing reliable predictions.

The performance of RUM-NN is rigorously evaluated in terms of parameter recovery and prediction accuracy, comparing it against the standard MNL and MNP specifications using both synthetic and real-world data. The proposed model, RUM-NN, can be designed in both linear and nonlinear structures. Across various scenarios, when the error terms in RUM-NN are set to IID Gumbel distributions, this model demonstrates accurate recovery of true parameters in synthesised datasets with 10,000 observations and 3 alternatives featuring IID Gumbel error terms. This closely matches the performance of MNL, its counterpart in the realm of econometrics. The parameter recovery was consistently accurate when error terms followed IID Normal distributions. In this scenario, RUM-NN demonstrated comparable performance to MNP, which is its counterpart for normal error terms. Furthermore, in scenarios with correlated error terms, RUM-NN could recover the correlation coefficient accurately.

To further evaluate RUM-NN, its performance was tested on two real-world datasets, the Swissmetro and LPMC datasets. On the Swissmetro dataset, the linear RUM-NN with Gumbel and Normal distributions replicated the performance of the MNL and MNP models, respectively, in terms of in-sample and out-sample log-likelihood values and prediction accuracy. For nonlinear structures, RUM-NN outperformed the DNN model, achieving higher log-likelihood and prediction accuracy. When the error distribution in RUM-NN was assumed to be Exponential, it achieved a significantly higher prediction accuracy of 76.93% compared to the MNL's 65.69%, the MNP's 66.85%, and the DNN's 70.21%. On the LPMC dataset, similar trends were observed. The linear RUM-NN with Gumbel and Normal distributions closely aligned with the performance of the MNL and MNP models, respectively. For nonlinear structures, RUM-NN again outperformed the DNN model in both prediction accuracy and log-likelihood. Among the tested error distributions, RUM-NN with an Exponential distribution achieved the highest prediction accuracy on the LPMC dataset, with a training accuracy of 67.92% and a testing accuracy of 67.54%, surpassing the DNN, MNL, and MNP models.

RUM-NN offers a flexible specification to accommodate any distributional form for the error terms and can capture correlation among error terms. The performance of this model can be further optimised under specific error term structures which are common in econometrics discrete choice methods. Improving the application and performance of RUM-NN for specific error term structures warrants further investigation.

## Appendix I

Table 8: Statistical analysis of linear RUM-NN with Gumbel distribution assumption and MNL for Swissmetro dataset

| Methods / Parameters | T-Test p-value | T-Test Conclusion | TOST p-value | TOST Conclusion |
|---|---|---|---|---|
| $ASC\_SM$ | 0.505 | Not different | 0.000 | Equivalent |
| $ASC\_Train$ | 0.729 | Not different | 0.000 | Equivalent |
| $\beta_{Cost}$ | 0.335 | Not different | 0.000 | Equivalent |
| $\beta_{Time}$ | 0.892 | Not different | 0.000 | Equivalent |
| $\beta_{Age\_Train2}$ | 0.781 | Not different | 0.000 | Equivalent |
| $\beta_{Age\_SM2}$ | 0.679 | Not different | 0.000 | Equivalent |
| $\beta_{Age\_Train3}$ | 0.965 | Not different | 0.000 | Equivalent |
| $\beta_{Age\_SM3}$ | 0.401 | Not different | 0.000 | Equivalent |
| $\beta_{Age\_Train4}$ | 0.539 | Not different | 0.000 | Equivalent |
| $\beta_{Age\_SM4}$ | 0.988 | Not different | 0.000 | Equivalent |
| $\beta_{Age\_Train5}$ | 0.659 | Not different | 0.000 | Equivalent |
| $\beta_{Age\_SM5}$ | 0.659 | Not different | 0.000 | Equivalent |
| $\beta_{Income\_Train2}$ | 0.439 | Not different | 0.000 | Equivalent |
| $\beta_{Income\_SM2}$ | 0.967 | Not different | 0.000 | Equivalent |
| $\beta_{Income\_Train3}$ | 0.3958 | Not different | 0.000 | Equivalent |
| $\beta_{Income\_SM3}$ | 0.945 | Not different | 0.000 | Equivalent |

Table 9: Statistical analysis of linear RUM-NN with Normal distribution assumption and MNL for LPMC dataset

| Methods / Parameters | T-Test p-value | T-Test Conclusion | TOST p-value | TOST Conclusion |
|---|---|---|---|---|
| $ASC\_CYCLING$ | 0.162 | Not different | 0.000 | Equivalent |
| $ASC\_PT$ | 0.101 | Not different | 0.000 | Equivalent |
| $ASC\_DRIVING$ | 0.866 | Not different | 0.000 | Equivalent |
| $\beta_{Time\_Walking}$ | 0.873 | Not different | 0.000 | Equivalent |
| $\beta_{Time\_Cycling}$ | 0.223 | Not different | 0.000 | Equivalent |
| $\beta_{Time\_Driving}$ | 0.376 | Not different | 0.000 | Equivalent |
| $\beta_{Cost\_Driving}$ | 0.926 | Not different | 0.000 | Equivalent |
| $\beta_{Cost\_PT}$ | 0.200 | Not different | 0.000 | Equivalent |
| $\beta_{Time\_PT\_BUS}$ | 0.124 | Not different | 0.000 | Equivalent |
| $\beta_{Time\_PT\_Rail}$ | 0.672 | Not different | 0.000 | Equivalent |
| $\beta_{Time\_PT\_Access}$ | 0.264 | Not different | 0.000 | Equivalent |
| $\beta_{Time\_PT\_INT\_Wait}$ | 0.569 | Not different | 0.000 | Equivalent |
| $\beta_{Traffic\_Driving}$ | 0.330 | Not different | 0.000 | Equivalent |
| $\beta_{Age\_Driving}$ | 0.999 | Not different | 0.000 | Equivalent |
| $\beta_{Age\_Walking}$ | 0.176 | Not different | 0.000 | Equivalent |
| $\beta_{Age\_Cycling}$ | 0.801 | Not different | 0.000 | Equivalent |
| $\beta_{Gender\_Driving}$ | 0.271 | Not different | 0.000 | Equivalent |
| $\beta_{Gender\_Walking}$ | 0.694 | Not different | 0.000 | Equivalent |
| $\beta_{Gender\_Cycling}$ | 0.820 | Not different | 0.000 | Equivalent |

31